\documentclass[10pt, conference]{IEEEtran}
\IEEEoverridecommandlockouts
\usepackage{cite}
\usepackage{makeidx} 
\usepackage{booktabs}
\usepackage{graphicx}
\usepackage[utf8]{inputenc}
\usepackage{csquotes}
\usepackage{amsmath,amssymb,amsfonts}
\usepackage{lscape}
\usepackage{pifont}
\usepackage{subcaption}
\usepackage{enumerate}
\usepackage{enumitem}
\usepackage{hyperref}
\usepackage{multirow}
\usepackage{todonotes}
\usepackage{algorithm}
\usepackage{algpseudocode}
\usepackage{soul}
\usepackage{listings}
\usepackage{pgfplots}
\usepackage{float}

\usepackage[framemethod=TikZ]{mdframed}

\newcommand{\bug}[1]{\texttt{#1}}

\begin{document}
%\mainmatter
\title{Patch Quality and Diversity of Invariant-Guided Search-Based Program Repair}

\author{\IEEEauthorblockN{Zhen Yu Ding}
	    \IEEEauthorblockA{\textit{University of Pittsburgh} \\
	    				Pittsburgh, PA, USA \\
	    				zhd23@pitt.edu}}

\maketitle

\begin{abstract}
Most automatic program repair techniques rely on test cases to specify correct program behavior.
Due to test cases' frequently incomplete coverage of desired behavior, however, patches often 
overfit and fail to generalize to broader requirements.
Moreover, in the absence of perfectly correct outputs, methods to promote higher patch quality, 
such as merging together several patches or a human evaluating patch recommendations, benefit from 
having access to a diverse set of patches, making patch diversity a potentially useful trait.
We evaluate the correctness and diversity of patches generated by GenProg and an invariant-based 
diversity-enhancing extension described in our prior work. We find no evidence that promoting 
diversity changes the correctness of patches in a positive or negative direction. Using invariant- and 
test case generation-driven metrics for measuring semantic diversity, we find no observed semantic 
differences between patches for most bugs, regardless of the repair technique used.
\end{abstract}

\section{Introduction}
Software bugs are troublesome: a study of software bugs in 2017 found that just 606 known 
bugs affected half of the population and \$1.7 trillion USD of assets~\cite{tricentis}.
They are also cumbersome for programmers, who spend, on average, approximately half of their 
time diagnosing and squashing bugs~\cite{cambridge-study}. Efforts in automated program repair 
attempt to alleviate these societal and developmental costs by automating the bug repair process.
Most program repair tools require the user to input a program and a set of test cases, which specify 
desired program behavior. Test cases, however, frequently fail to fully cover the actual specifications 
of a program. Consequently, program repair techniques frequently produce incorrect patches that 
overfit to the provided test suite and fail to generalize to broader requirements not explicitly specified 
in the provided tests~\cite{patch-correctness, d4j-eval, Smith15fse, Le2018}. Producing correct patches 
that adhere to broader, implicit specifications is thus a desideratum of program repair techniques.

Inconsistency in the quality of patches generated by automated program repair tools can perhaps 
be alleviated with, among other means, a human in the loop judging patches and/or by combining information from 
patches to produce a superior patch~\cite{mau-proposal}. In both cases, providing a set of diverse patches may be 
useful. For the human in the loop adjudicating patches, a diversity of patches means a diversity of 
options to choose from. For the approach of merging patches into a better patch, a diversity of 
patches means a diversity of information to merge together. Producing diverse patches may thus also 
be desirable for these applications.

We thus evaluate patch correctness and diversity for two program repair techniques: GenProg~\cite{genprog}, 
a search-based program repair technique based on genetic programming, and a invariant-driven diversity-promoting 
repair technique derived from GenProg and described in our prior work~\cite{dinglyu}. 
To evaluate patch correctness, we use held-out tests, a practice previously used in evaluating 
program repair tools~\cite{Smith15fse, Le2018}. To evaluate patch diversity, we use semantic diversity 
metrics based on inferred invariants~\cite{dinglyu} and automatic test case generation tools~\cite{mau-proposal}.

This paper's main contributions are: 
\begin{itemize}
	\item An evaluation of patch correctness for GenProg~\cite{genprog} and our diversity-promoting repair technique~\cite{dinglyu}.
	\item An evaluation of patch diversity using both invariant-based~\cite{dinglyu} 
	and test generation-based~\cite{mau-proposal} semantic diversity for GenProg and 
	our diversity-promoting repair technique.
\end{itemize}

\section{Background}
\label{sec:background}

\subsection{GenProg and Attempts at Improvement}
GenProg~\cite{genprog} is a search-based automated program repair 
technique. GenProg accepts, as input, a faulty program's 
source code, failing test cases (negative tests) that demonstrate the fault, and passing 
test cases (positive tests) that demonstrate desirable program behavior to preserve. 
To produce a bug patch, GenProg treats bug repair as an optimization problem: 
which patch(es) in the space of possible patches passes as many test cases as possible?
The globally optimal solutions are patches that pass all test cases, which GenProg assumes 
to be correct. GenProg solves this optimization problem using genetic programming.
To traverse the space of patches, GenProg mutates and recombines code at the statement level, 
creating a population of variant programs, or \emph{candidate patches}.
In contrast to approaches that randomly traverse the space of patches, such as TrpAutoRepair~\cite{trp}, 
GenProg guides the traversal by observing the number of test cases that each candidate patch 
passes. For each candidate patch, GenProg computes a fitness score equal to a weighted sum of the 
number of positive and negative patches that the candidate patch passes. Candidate patches 
with higher fitness are more likely to reproduce through further mutation and recombination, and 
their offspring will carry the code edits of the parent candidate patch, unless if negated by further 
code edits.

GenProg's reliance on test cases as a measurement of program correctness makes 
the approach vulnerable to overfitting, where GenProg constructs a patch that conforms to the 
test suite, but breaks other implicit software requirements not covered by the 
provided tests. The weakness of test cases as a specification for program behavior is 
not specific to GenProg; the problem is common among program repair techniques that 
rely on test cases as a metric of program correctness.
Qi et al.~\cite{patch-correctness} found that almost all repairs produced by 
GenProg, AE~\cite{ae}, and RSRepair~\cite{rsrepair}---techniques which depend on 
test cases to specify desired behavior---are incorrect. They also find that many patches introduce 
undesirable effects, such as deleting functionality or introducing security vulnerabilities.
Other studies, which we discuss in more detail in Section~\ref{sec:relwork}, have also found a high proportion 
of incorrect patches among various 
program repair techniques~\cite{d4j-eval, Smith15fse, Le2018}.

Moreover, GenProg's test case-based fitness function should, in theory,
identify and favor partial solutions that pass more test cases than the original program. 
In practice, however, the intuition fails to hold. Instead, distinct candidate patches often 
receive the same fitness score, resulting in a fitness landscape with large, flat plateaus~\cite{source-code-checkpoint, gecco09}.
In the common case where a bug is revealed by only a single negative test, there may 
not exist any intermediate fitness values between the fitness of the original program and 
a ``correct" patch. The fitness function's inability to differentiate between candidate patches 
undermines GenProg's ability to compose more complex patches. Despite being 
capable of constructing multi-edit repairs, much of GenProg's patches actually reduce down to 
single-edit repairs~\cite{patch-correctness}. A potential explanation of test cases' 
weakness at discriminating candidate patches may lie in the binary nature of test cases. 
Test cases generally only pass or fail, and generally do not provide indications of 
partial correctness. As a result, test cases lack granularity in the semantic information 
they provide, and a test case-based fitness function would thus also be not granular. 
Thus, a test suite might determine different patches with differing levels of partial correctness 
to be equally incorrect.

Our previous work~\cite{dinglyu} attempts to improve on GenProg's search strategy by 
incorporating additional semantic information from inferred invariants. 
Using semantic information derived from invariants, we determine which patches 
appear to be more semantically unique and diverse, which we favor for selection and 
further exploration. By leveraging added semantic information to diversify the search 
of the repair space, we sought to discourage the search process from being trapped 
in locally optimal fitness values, which are undesirable in genetic algorithms~\cite{intro-to-ec}.
Moreover, producing multiple semantically diverse patches may be desirable for 
patch recommendation systems, where a human in the loop may choose between 
several semantically diverse patches, or for combining multiple low quality patches 
into a higher quality patch~\cite{mau-proposal}.
We describe the approach's method of measuring and promoting semantic diversity 
in~\ref{sec:invdiv}. 
An evaluation on a refactored subset of IntroClassJava~\cite{introclassjava} reveals that 
while our prior work was able to successfully diversify the search 
for a repair and better differentiate between candidate patches, we had no statistically 
significant evidence of an improvement in repair efficiency or success.

There exist other previous efforts to incorporate more semantic information to improve 
GenProg's search process. Fast et al.~\cite{better-fitness}
attempts to learn correlations between invariant behavior and program success or failure. 
Their approach uses known patches for a bug to train a linear model that maps the behavior 
of invariants and test cases to fitness values. de Souza et al.~\cite{source-code-checkpoint} 
uses dynamic analysis to track changes in variable values at pre-determined source code 
checkpoints. They produce a checkpoints metric based on these changes, which they 
incorporate into the fitness function.

\subsection{Daikon and Invariant-Based Semantic Diversity}
\label{sec:invdiv}
Daikon~\cite{daikon} is a dynamic analysis tool that infers likely program invariants. 
Program invariants are properties which hold true at indicated program locations. 
Given a program and a method to execute the program (such as test cases), Daikon collects 
execution traces and analyzes which invariants, as described by a grammar, are statistically 
likely to be true for the program, as evidenced by program values in the collected trace.

Our previous work~\cite{dinglyu} introduces a measurement of semantic difference based 
on inferred invariants in the context of program repair. 
Given a set of likely invariants inferred from the original faulty program, we 
construct an \emph{invariant profile}---a string representation of the reachability and truth of 
the set of invariants in a program---for each candidate patch. 
We construct this invariant profile by instrumenting candidate patches to detect whether each invariant 
is reached and ever violated during program execution, with the results recorded separately for 
executions of positive and negative tests.
The semantic distance between two programs is the Hamming distance between the two 
programs' string-based invariant profiles. The semantic diversity of a program within a 
population of programs is the semantic distance between the program and the rest of the 
population, which we compute by summing the semantic distance between the program and 
each other program in the population. We promote semantic diversity by using our semantic 
diversity metric as an optimization objective along with test cases.

\subsection{EvoSuite and Test Generation-Based Semantic Diversity}
EvoSuite~\cite{evosuite} is a search-based automatic test case generation tool for Java programs.
Given a Java bytecode program, EvoSuite produces JUnit tests~\cite{evosuite-ssbse18-tutorial}.
To generate tests, EvoSuite uses a combination of mutation analysis and concolic execution 
to produce and evolve a test suite that maximizes code coverage. Mutating the program simulates the 
introduction of faults, and EvoSuite uses genetic programming to create tests that detect these mutations, 
with concolic execution activated after a preset number of generations.

Soto~\cite{mau-proposal} proposes a measurement for semantic distance based on the output of 
test case generators. Given two programs $P$ and $Q$ to compare, Soto uses EvoSuite to generate 
test suites $T_P$ and $T_Q$ and merges them to produce $T_P \cup T_Q$. To describe the semantics 
of $P$ and $Q$, Soto evaluates each program on the merged test suite $T_P \cup T_Q$ to produce a 
string-based report. The semantic distance between $P$ and $Q$ is the Hamming distance between the 
two programs' reports, normalized by the length of the reports ($|T_P \cup T_Q|$).

\section{Analysis}
\label{sec:analysis}
We evaluate the correctness and semantic diversity of patches generated by GenProg~\cite{genprog}
and the program repair approach described by our prior work~\cite{dinglyu}. We address the following 
research questions:
\begin{description}
\item [RQ1:] \textbf{Does promoting semantic diversity using invariants change the correctness of repairs generated?}

Since both repair approaches rely on test cases to specify desired program behavior, both techniques 
risk creating patches that overfit to the provided tests and fail to generalize to broader, implicit requirements.
Previous evaluation of GenProg found that many of its repairs are, in fact, incorrect due to weak test suites~\cite{patch-correctness}. 
Promoting semantic diversity may inadvertently encourage an accumulation of negative, fault-creating 
edits that test cases might fail to detect, since these edits may result in a change in invariant-observed 
semantic behavior and thus be favored. Conversely, if truly correct patches are sparser in semantic space 
than falsely correct patches, then broadening and diversifying the search throughout semantic space 
might result in a higher chance of encountering truly correct repairs. We thus investigate patch correctness 
for both repair techniques.

\item [RQ2:] \textbf{Does promoting semantic diversity using invariants change the semantic diversity of repairs generated?}

By encouraging a broad and diverse search through semantic space, we hypothesize that our previously 
described approach would produce semantically diverse patches. A diverse set of patches may be 
useful for applications with a human in the loop, such as patch recommendation systems, or for combining 
multiple patches into a better patch~\cite{mau-proposal}. We thus evaluate patch diversity for both repair techniques.
\end{description}

To evaluate our research questions, we use the modified version of IntroClassJava~\cite{introclassjava} created 
for our prior work~\cite{dinglyu}. IntroClassJava is a derivative of IntroClass~\cite{manybugs}, 
a set of buggy programs written by students in an introductory programming class. Every IntroClassJava bug 
comes with a set of black-box tests and white-box tests. Black-box tests derive from the program's 
specifications and are written manually by the class's instructor. White-box tests derive from 
generating tests using KLEE~\cite{klee}, a symbolic execution-based test generator, on a C reference 
implementation, with manually written tests to cover branches that KLEE cannot generate a test for. 
We use white-box tests as input to GenProg and our diversity-promoting repair technique, and we use 
black-box tests as a held-out test suite to evaluate the correctness of patches generated by both techniques
as part of RQ1. Note that this evaluation setup differs from the setup used in evaluating our prior work~\cite{dinglyu}, 
where we inputted both white-box and black-box tests to both repair techniques.

We discard any patches found during the initialization process of either technique, as GenProg and our 
diversity-promoting technique differ only after the initialization process, where selective pressure begins 
to shape the search for a repair. As a result, both techniques traverses the search space identically 
during initialization, and any patches found during initialization do not illustrate the differences in behavior 
of the two techniques. Therefore, we exclude such patches from analysis.

For our preliminary experiments described here,
we use four bugs from the six bugs successfully repaired by either repair technique in our prior work~\cite{dinglyu}. 
We use a subset of our previously repaired bugs---which pass all white-box and black-box tests---in
order to guarantee that there exists at least one ``correct" patch that passes both sets of tests within the 
search space of the techniques. Out of the six bugs, we exclude \bug{smallext-af81-000} as the bug 
does not have any failing white-box tests, and we exclude \bug{smallest-8839-002} as both techniques 
found patches exclusively during the initialization process in our testing. We thus use the four remaining bugs.

For both techniques, we use 
the append, replace, and delete mutation operators, one-point crossover, tournament selection with 
a tournament size of $k=2$, a population size of 40, and a limit of 10 generations. We repeat each 
experiment with random seeds 0--9 inclusive. We conduct our experiments on a computer with the 
following specifications: Ubuntu Server 16.04 LTS, 4x Intel Xeon E7-4820, 128 GB of RAM, 
and 2 TB of magnetic hard disk storage.

\subsection{Evaluation of RQ1: Patch Correctness}

\begin{table}[t]
	\begin{center}
		\begin{tabular}{lrr}
			\toprule
			\textbf{Bug} & \multicolumn{2}{c}{\textbf{Number of ``Correct" Patches}} \\
			& \textbf{GenProg} & \textbf{Ding et al.}\\
			\midrule
			\bug{digits-0cdf-006} & 1 / 6 & 1 / 7 \\
			\bug{digits-0cdf-007} & 4 / 4 & 4 / 4 \\
			\bug{digits-5b50-000} & 0 / 7 & 0 / 7 \\
			\bug{digits-d120-001} & 7 / 8 & 8 / 8 \\
			\midrule
			Total & 12 / 25 & 13 / 26\\
			\bottomrule
		\end{tabular}
	\end{center}
		\caption{Number of ``correct" patches generated by each repair technique compared to 
		the number of patches generated after initialization. 
		A patch is deemed 
		``correct" if it passes all tests in the held-out black-box test suite. All patches deemed incorrect 
		fail exactly one black-box test.}
		\label{table:correctness}
\end{table}

We evaluate patch correctness by testing each patch against its held-out black-box tests.
Table~\ref{table:correctness} provides the proportion of ``correct" patches generated by each technique. 
The proportion of patches that the black-box tests deem correct are almost identical for both techniques.
A two-sided Fisher's exact test reveals no statistically significant evidence ($p > 0.05$) of a difference in 
the number of correct and incorrect patches generated by both techniques. Moreover, most bugs demonstrate 
all-or-nothing behavior, where almost all repairs generated by either technique are correct or incorrect, 
suggesting that the strength of test suites takes precedence over the search strategy in producing correct 
patches.

\subsection{Evaluation of RQ2: Patch Diversity}

\begin{table}[t]
	\begin{center}
		\begin{tabular}{lrrrr}
			\toprule
			\textbf{Bug} & \multicolumn{4}{c}{\textbf{Mean Semantic Diversity}} \\
			& \multicolumn{2}{c}{\textbf{Invariant-Based}} & \multicolumn{2}{c}{\textbf{EvoSuite-Based}} \\
			& \textbf{GenProg} & \textbf{Ding et al.} & \textbf{GenProg} & \textbf{Ding et al.}\\
			\midrule
			\bug{digits-0cdf-006} & 0 & 0 & 0 & 0 \\
			\bug{digits-0cdf-007} & 0 & 0 & 0 & 0 \\
			\bug{digits-5b50-000} & 0 & 0 & 0.5 & 0.08 \\
			\bug{digits-d120-001} & 0 & 0 & 0 & 0 \\
			\bottomrule
		\end{tabular}
	\end{center}
		\caption{Average semantic diversity of patches generated by each repair technique, 
		measured by both inferred invariants~\cite{dinglyu} and EvoSuite-generated tests~\cite{mau-proposal}.}
		\label{table:sem-div}
\end{table}

\begin{table}[t]
	\begin{center}
		\begin{tabular}{lrr}
			\toprule
		\end{tabular}
	\end{center}
\end{table}

We evaluate patch diversity using both invariant-based and test generation-based metrics of 
semantic diversity. To measure invariant-based diversity, we treated the set of repairs for a particular bug 
as a population and used the metric for invariant-based semantic diversity described in our prior work~\cite{dinglyu}. 
To measure test generation-based semantic diversity, we used the metric for semantic distance described 
by Soto~\cite{mau-proposal}, and summed the semantic distance between a patch and every other patch for a 
particular bug to compute a patch's semantic diversity. Table~\ref{table:sem-div} provides the average 
semantic diversity of patches generated by both techniques. For both techniques and almost all bugs, 
neither metric for semantic diversity discerned any semantic difference between any patches for a particular bug.

The semantic indistinguishability of patches suggests either a lack of patch diversity or weakness in 
the two metrics of semantic distance. We intend to conduct a manual patch analysis to investigate 
patch semantics and determine the cause of indistinguishability in future work.

\section{Threats to Validity}
\label{sec:threats}

\subsection{Test Case Weakness}
Using held-out test cases to evaluate patch correctness poses the risk of a patch overfitting to both 
the white-box and black-box tests. 
While patches deemed correct by the held-out tests may still be 
incorrect, using held-out tests nevertheless allows us to determine whether patches generalize to requirements 
not specified in the inputted tests but evaluated in the held-out tests. Manual correctness analysis and 
formal verification may provide stronger evidence of correctness.

\subsection{Sample Size}
Our early results evaluate only four bugs and 51 patches. The four bugs are chosen as test 
subjects as there exist known patches within the search space that are correct with respect to both white-box and black-box 
tests. We propose to broaden our evaluation to more software bugs in future work.

\subsection{Bug Benchmark}
We use a subset of IntroClassJava~\cite{introclassjava} programs as a benchmark for evaluation. 
These buggy programs are small ($< 30$ LOC) and are written by introductory programming 
students as part of class assignments, rather than by professional software engineers as a part of 
real-world software. Moreover, the bugs' test suites were designed to have very high specification 
coverage (with black-box tests) or branch coverage (with white-box tests)~\cite{manybugs}, which 
is not always true in real-world software.
Evaluating IntroClassJava programs and their bugs does not simulate an application of program repair
on industrially-sized software. We propose to evaluate patch correctness and diversity on benchmark 
datasets containing larger, real-world programs, such as Defects4J~\cite{defects4j} or Bugs.Jar~\cite{bugsdotjar}.

\section{Related Work}
\label{sec:relwork}
This section discusses other prior efforts in evaluating patch quality. We are not aware of prior 
efforts in evaluating patch diversity at the time of writing.

Qi et al.~\cite{patch-correctness} evaluates the correctness of patches generated by the 
search-based program repair tools GenProg~\cite{genprog}, AE~\cite{ae}, and RSRepair~\cite{rsrepair}.
They find that the vast majority of patches generated by these tools are incorrect. When they 
write additional test cases to expose the errors found and input them into GenProg to better guide 
the search, GenProg was unable to find any patches. They also found that most patches generated 
by all three techniques are semantically reducible to functionality deletion and often introduce undesired 
effects, such as security flaws or loss of functionality.

Martinez et al.~\cite{d4j-eval} evaluates the correctness of patches that GenProg generates 
for the Defects4J~\cite{defects4j} benchmark, a set of real-world Java bugs in open source software.
Manual analysis reveals that a large majority of patches were incorrect.

Smith et al.~\cite{Smith15fse} evaluates the correctness of patches generated by the 
search-based program repair tools GenProg~\cite{genprog} and TrpAutoRepair~\cite{trp}, 
as well as compare the correctness of tool-generated patches to patches written by novice student programmers.
They use IntroClass as a benchmark and also use held-out tests to evaluate patch correctness 
in a manner similar to our evaluation. They find that a large majority of patches are incorrect based 
on the held-out tests, but also that there's no statistically significant difference between the correctness 
rate of tool and novice human-generated tests.

Le et al.~\cite{Le2018} evaluates the correctness of patches generated by semantics-based program 
repair techniques, which use symbolic execution and program synthesis, rather than a search through 
the space of possible patches, to generate a repair. They evaluate the semantics-based repair tool 
Angelix~\cite{angelix} and variations of the tool with different synthesis engines~\cite{sygus-repair} on 
the IntroClass~\cite{manybugs} and Codeflaws~\cite{codeflaws} benchmarks, both of which supply two separate 
test suites per bug. Similar to our evaluation of patch correctness, they use one of the two independent 
test suites as held-out tests for evaluating patch correctness. They find that a large majority of patches 
generated by all techniques are incorrect based on the held-out tests, revealing that the problem of 
patch overfitting extends to semantic-based repair techniques as well.

\iffalse
Le et al.~\cite{le-patch-correctness-eval} evaluates the reliability of using held-out tests and 
manual analysis as means of determining patch correctness. They conduct a user study with 
professional software developers construct a gold-standard set of patch correctness labels.
\fi

Xin and Reiss~\cite{difftgen} proposes DiffTGen, a technique for evaluating patch correctness by 
generating test cases based on differences in program syntax between a buggy program and its 
patch. If running these generated tests produces different output, the difference is sent to an 
oracle (a human in their evaluation) for correctness analysis. DiffTGen identified 
49.4\% of overfitting patches generated by various program repair techniques. Their tool can also 
enhance automated program repair techniques by augmenting test suites with additional generated 
tests.

\section{Conclusion}
\label{sec:conclusion}
Our initial results do not suggest evidence of diversity promotion causing an increase or decrease in 
patch correctness. We moreover find that our metrics for semantic diversity were unable to distinguish 
between patches for almost every bug, regardless of repair technique used. We propose to further 
investigate the cause of semantic patch indistinguishability with further analyses, including manual 
analysis. Moreover, we intend to broaden the scope of bugs for investigation, including real-world bugs 
in larger, industrially-sized software.

\bibliographystyle{IEEEtran}
\bibliography{IEEEabrv,references}

\begin{thebibliography}{10}
\providecommand{\url}[1]{#1}
\csname url@samestyle\endcsname
\providecommand{\newblock}{\relax}
\providecommand{\bibinfo}[2]{#2}
\providecommand{\BIBentrySTDinterwordspacing}{\spaceskip=0pt\relax}
\providecommand{\BIBentryALTinterwordstretchfactor}{4}
\providecommand{\BIBentryALTinterwordspacing}{\spaceskip=\fontdimen2\font plus
\BIBentryALTinterwordstretchfactor\fontdimen3\font minus
  \fontdimen4\font\relax}
\providecommand{\BIBforeignlanguage}[2]{{%
\expandafter\ifx\csname l@#1\endcsname\relax
\typeout{** WARNING: IEEEtran.bst: No hyphenation pattern has been}%
\typeout{** loaded for the language `#1'. Using the pattern for}%
\typeout{** the default language instead.}%
\else
\language=\csname l@#1\endcsname
\fi
#2}}
\providecommand{\BIBdecl}{\relax}
\BIBdecl

\bibitem{tricentis}
``Software fail watch: 5th edition,''
  \url{https://www.tricentis.com/software-fail-watch/}, {Accessed December,
  2018}.

\bibitem{cambridge-study}
``Cambridge university study states software bugs cost economy {\$}312 billion
  per year,'' \url{http://www.prweb.com/releases/2013/1/prweb10298185.htm},
  {Accessed December, 2018}.

\bibitem{patch-correctness}
Z.~Qi, F.~Long, S.~Achour, and M.~Rinard, ``An analysis of patch plausibility
  and correctness for generate-and-validate patch generation systems,'' in
  \emph{International Symposium on Software Testing and Analysis}, ser. ISSTA
  '15, 2015, pp. 24--36.

\bibitem{d4j-eval}
M.~Martinez, T.~Durieux, R.~Sommerard, J.~Xuan, and M.~Monperrus, ``Automatic
  repair of real bugs in java: A large-scale experiment on the defects4j
  dataset,'' \emph{{Empirical Software Engineering}}, vol.~22, no.~4, pp.
  1936--1964, 2017.

\bibitem{Smith15fse}
E.~K. Smith, E.~Barr, C.~Le~Goues, and Y.~Brun, ``Is the cure worse than the
  disease? {Overfitting} in automated program repair,'' in \emph{Joint Meeting
  of the European Software Engineering Conference and the Symposium on the
  Foundations of Software Engineering}, 2015, pp. 532--543.

\bibitem{Le2018}
X.~D. Le, F.~Thung, D.~Lo, and C.~Le~Goues, ``Overfitting in semantics-based
  automated program repair,'' \emph{Empirical Software Engineering}, vol.~23,
  no.~5, pp. 3007--3033, 2018.

\bibitem{mau-proposal}
M.~Soto, ``Improving patch quality by enhancing key components of automatic
  program repair,'' Ph.D. dissertation proposal, Carnegie Mellon University,
  April 2019.

\bibitem{genprog}
C.~Le~Goues, T.~Nguyen, S.~Forrest, and W.~Weimer, ``Gen{P}rog: A generic
  method for automatic software repair,'' \emph{IEEE Trans. Software Eng.},
  vol.~38, pp. 54--72, 2012.

\bibitem{dinglyu}
Z.~Ding, Y.~Lyu, C.~S. Timperley, and C.~{Le~Goues}, ``Leveraging program
  invariants to promote population diversity in search-based automatic program
  repair,'' in \emph{Genetic Improvement Workshop}, 2019, p. (to appear).

\bibitem{trp}
Y.~Qi, X.~Mao, and Y.~Lei, ``{Efficient Automated Program Repair through
  {Fault-Recorded} Testing Prioritization},'' in \emph{International Conference
  on Software Maintenance}, 2013, pp. 180--189.

\bibitem{ae}
W.~Weimer, Z.~P. Fry, and S.~Forrest, ``{Leveraging program equivalence for
  adaptive program repair: Models and first results},'' in \emph{International
  Conference on Automated Software Engineering}, ser. ASE '13, 2013, pp.
  356--366.

\bibitem{rsrepair}
Y.~Qi, X.~Mao, Y.~Lei, Z.~Dai, and C.~Wang, ``The strength of random search on
  automated program repair,'' in \emph{International Conference on Software
  Engineering}, ser. ICSE '14, 2014, pp. 254--265.

\bibitem{source-code-checkpoint}
E.~F. de~Souza, C.~Le~Goues, and C.~G. Camilo-Junior, ``A novel fitness
  function for automated program repair based on source code checkpoints,'' in
  \emph{Genetic and Evolutionary Computation Conference}, ser. GECCO '18, 2018.

\bibitem{gecco09}
S.~Forrest, W.~Weimer, T.~Nguyen, and C.~Le~Goues, ``A genetic programming
  approach to automated software repair,'' in \emph{Genetic and Evolutionary
  Computation Conference (GECCO)}, 2009, pp. 947--954.

\bibitem{intro-to-ec}
A.~E. Eiben and J.~E. Smith, \emph{Introduction to evolutionary
  computing}.\hskip 1em plus 0.5em minus 0.4em\relax Springer, 2011.

\bibitem{introclassjava}
T.~Durieux and M.~Monperrus, ``{IntroClassJava}: {A} benchmark of 297 small and
  buggy {Java} programs,'' {Universite Lille 1}, Research Report, 2016.

\bibitem{better-fitness}
E.~Fast, C.~Le~Goues, S.~Forrest, and W.~Weimer, ``Designing better fitness
  functions for automated program repair,'' in \emph{Genetic and Evolutionary
  Computation Conference}, ser. GECCO '10, 2010, pp. 965--972.

\bibitem{daikon}
M.~D. Ernst, J.~H. Perkins, P.~J. Guo, S.~McCamant, C.~Pacheco, M.~S. Tschantz,
  and C.~Xiao, ``{The {Daikon} system for dynamic detection of likely
  invariants},'' \emph{Science of Computer Programming}, vol.~69, no. 1--3, pp.
  35--45, 2007.

\bibitem{evosuite}
G.~Fraser and A.~Arcuri, ``Evo{S}uite: automatic test suite generation for
  object-oriented software,'' pp. 416--419, 2011.

\bibitem{evosuite-ssbse18-tutorial}
G.~Fraser, ``A tutorial on using and extending the {EvoSuite} search-based test
  generator,'' in \emph{Search-Based Software Engineering}.\hskip 1em plus
  0.5em minus 0.4em\relax Springer, 2018, pp. 106--130.

\bibitem{manybugs}
C.~Le~Goues, N.~Holtschulte, E.~K. Smith, Y.~Brun, P.~Devanbu, S.~Forrest, and
  W.~Weimer, ``The {ManyBugs} and {IntroClass} benchmarks for automated repair
  of {C} programs,'' \emph{IEEE Trans. Software Eng.}, vol.~41, no.~12, pp.
  1236--1256, 2015.

\bibitem{klee}
C.~Cadar, D.~Dunbar, and D.~Engler, ``{KLEE}: {U}nassisted and automatic
  generation of high-coverage tests for complex systems programs,'' in
  \emph{{USENIX} Symposium on Operating Systems Design and Implementation},
  ser. OSDI '08, 2008.

\bibitem{defects4j}
R.~Just, D.~Jalali, and M.~D. Ernst, ``{Defects4J}: A database of existing
  faults to enable controlled testing studies for {J}ava programs,'' in
  \emph{International Symposium on Software Testing and Analysis (ISSTA)},
  2014, pp. 437--440.

\bibitem{bugsdotjar}
R.~K. Saha, Y.~Lyu, W.~Lam, H.~Yoshida, and M.~R. Prasad, ``{Bugs.Jar}: {A}
  large-scale, diverse dataset of real-world {J}ava bugs,'' in
  \emph{International Conference on Mining Software Repositories}, ser. MSR
  '18, 2018, pp. 10--13.

\bibitem{angelix}
S.~Mechtaev, J.~Yi, and A.~Roychoudhury, ``Angelix: {S}calable multiline
  program patch synthesis via symbolic analysis,'' in \emph{International
  Conference on Software Engineering}, ser. ICSE '16, 2016, pp. 691--701.

\bibitem{sygus-repair}
X.~D. Le, D.~Lo, and C.~{Le Goues}, ``Empirical study on synthesis engines for
  semantics-based program repair,'' in \emph{International Conference on
  Software Maintenance and Evolution}, ser. ICSME '16, 2016, pp. 423--427.

\bibitem{codeflaws}
S.~H. Tan, J.~Yi, Yulis, S.~Mechtaev, and A.~Roychoudhury, ``Codeflaws: a
  programming competition benchmark for evaluating automated program repair
  tools,'' in \emph{International Conference on Software Engineering}, ser.
  ICSE '18, 2017.

\bibitem{difftgen}
Q.~Xin and S.~P. Reiss, ``Identifying test-suite-overfitted patches through
  test case generation,'' in \emph{International Symposium on Software Testing
  and Analysis}, ser. ISSTA '17, 2017.

\end{thebibliography}

\end{document}